\documentclass[aps,prb,twocolumn,superscriptaddress,showpacs,floatfix]{revtex4}
\usepackage{graphicx}
\begin{document}
\title{IR Phonon signatures of multiferroicity in TbMn$_{2}$O$_{5}$}
\author{R. Vald\'{e}s Aguilar}
\affiliation{Materials Research Science and Engineering Center,
University of Maryland, College Park, Maryland 20742}
\author{A. B. Sushkov}
\affiliation{Materials Research Science and Engineering Center,
University of Maryland, College Park, Maryland 20742}
\author{S. Park}
\affiliation{Rutgers Center for Emergent Materials and Department
of Physics and Astronomy, Rutgers University, Piscataway, New
Jersey 08854}
\author{S-W. Cheong}
\affiliation{Rutgers Center for Emergent Materials and Department
of Physics and Astronomy, Rutgers University, Piscataway, New
Jersey 08854}
\author{H. D. Drew}
\affiliation{Materials Research Science and Engineering Center,
University of Maryland, College Park, Maryland 20742}

\begin{abstract}
The infrared (IR) active phonons in multiferroic TbMn$_2$O$_5$ are
studied as a function of temperature. Most of the symmetry allowed
IR modes polarized along the \textit{a} and \textit{b} crystal
axes are observed and the behavior of several \textit{b} polarized
phonons is correlated with the magnetic and ferroelectric
transitions. A high frequency \textit{b} polarized phonon only
Raman allowed in the paraelectric phase becomes IR active at the
ferroelectric transition. The IR strength of this mode is
proportional to the square of the ferroelectric order parameter
and gives a sensitive measure of the symmetry lowering lattice
distortions in the ferroelectric phase.
\end{abstract}


\maketitle
\section{Introduction}
Materials that exhibit simultaneously (anti)-ferromagnetic,
(anti)-ferroelectric or (anti)-ferroelastic degrees of freedom,
the so called multiferroics, have become of increasing interest
because they offer great possibilities for applications in
multifunctional devices. These applications are specially based on
the power to control the magnetic state by electric fields and the
ferroelectric state by magnetic fields, this type of materials
being the magnetoelectrics, a subclass of the multiferroics
\cite{Hill-why,Fiebig-magnetoelectric-review,Prellier-review,Binek-appl}.
Moreover, the understanding of the basic mechanism that allows
this behavior is important in the development of the devices that
make use of these characteristics.

The interplay of these degrees of freedom has been demonstrated
recently in several multiferroic materials such as
\emph{R}Mn$_2$O$_5$ where \emph{R} is Eu, Gd, Tb, Dy, Ho and Y
\cite{Popov-YMn2O5,Hur-nature,Dy-Hur,Gd-Eu-Golovenchits,Higashiyama-Dy}
as well as in TbMnO$_3$ \cite{Kimura-113} and Ni$_3$V$_2$O$_8$
\cite{Lawes-vanadate}. Fundamental questions as to the nature and
characteristics of the coupling of the magnetic and ferroelectric
orders remain unanswered. Phenomenological approaches
\cite{Mostovoy-spiral,Brooks-incommensurate} based on the Landau
theory of phase transitions have been used to relate the symmetry of
the spin order with the appearance of ferroelectricity. The basic
conclusion of these proposals is that the non-collinearity of the
spin system is crucial in the development of ferroelectric order.
More microscopic studies \cite{Katsura-spin-current,Sergienko-DM}
have also pointed to the importance of non-collinearity of the spin
system.

The antiferromagnetic manganese oxide TbMn$_2$O$_5$ (orthorhombic
space group Pbam \# 55, Z = 4) is a multiferroic with
non-collinear magnetic order \cite{Chapon-str-anomalies} that
shows a strong magnetoelectric coupling effect \cite{Hur-nature}:
an applied magnetic field along the \emph{a} axis changes the sign
of the electrical polarization present along the \emph{b} axis.
The dielectric constant $\varepsilon$ along the \emph{b} axis has
anomalies associated with the distinct phase transitions at low
temperatures (see figure 1 in Hur, et al \cite{Hur-nature}): at
the N\'{e}el temperature T$_N$ $\approx$ 42 K no anomalies are
present in $\varepsilon$. There is a paraelectric to ferroelectric
phase transition at T$_C$ $\approx$ 38 K evident by a peak in the
dielectric constant. The magnetic order then locks in to a
commensurate structure (CM) with wave vector (1/2,0,1/4). At T
$\approx$ 24 K the magnetic order transforms into an
incommensurate structure (ICM) with a step-like feature in
$\varepsilon$; this anomaly is also accompanied by hysteresis
\cite{Hur-nature}. We note that the dielectric constants along the
\emph{a} and \emph{c} axis show no significant effects at these
phase transitions.

In EuMn$_2$O$_5$ Polyakov, et al \cite{Polyakov-neutron} found
displacements of the Mn$^{3+}$ ion along the \emph{a} axis at the
ferroelectric transition. They suggested that this behavior leads
to a change in symmetry from the space group Pbam to the
non-centrosymmetric group P2$_1$am (\# 26). In a analogous work on
the compound YMn$_2$O$_5$ \cite{Kagomiya-xray}, Kagomiya, et al
proposed similar displacements at the ferroelectric transition.
The displacements reported are very small ($\approx 0.007$ {\AA}),
which hints to a different origin of the ferroelectricity when
compared to the typical proper ferroelectrics. Other structural
works \cite{Chapon-str-anomalies,Blake-spin-str,Chapon-YMn2O5}
have not reported any signature of atomic displacements at the
ferroelectric phase transition in this family of compounds.
Moreover, Mihailova, et al \cite{Mihailova-Raman} reported a Raman
study in HoMn$_2$O$_5$ and TbMn$_2$O$_5$ and Garc\'{i}a-Flores, et
al \cite{Garcia-Flores-Raman} in BiMn$_2$O$_5$, EuMn$_2$O$_5$ and
DyMn$_2$O$_5$ as a function of temperature and found no evidence
of anomalous behavior of the Raman active phonons at the
ferroelectric transition temperature. Nevertheless, the existence
of a macroscopic dipole moment is evidence of the lack of
inversion symmetry in the FE phase, and motivates the study of the
dynamics of the lattice to look for further information about the
structural changes.

\begin{figure}[h t]
\includegraphics[width=1\columnwidth]{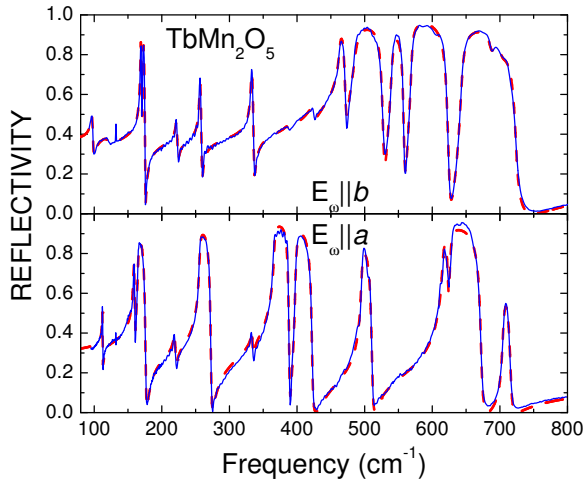}
\caption{(Color online). Experimental (solid line) and fit (dashed
line) at T = 7 K phonon spectra in TbMn$_2$O$_5$.}
\label{fullspectra}
\end{figure}

In this report we present a study of the temperature dependent
infrared (IR) phonon spectra of the multiferroic TbMn$_2$O$_5$. The
most interesting feature we find is the appearance of an IR inactive
phonon activated at the ferroelectric transition with light
polarization parallel to the static ferroelectric polarization
$\mathbf{P_0}$ ($E_\omega || \mathbf{P_0} || b$). This indicates
that one IR forbidden mode (Raman or silent) in the paraelectric
phase acquires an electric-dipole moment due to the static
displacement associated with the ferroelectricity. We identify this
phonon with a Raman Mn-O stretching mode, which accounts for its
sensitivity to the static polarization possibly induced by the Mn
spin system. \cite{Chapon-YMn2O5, Mostovoy-spiral}.

\section{Experimental Results}
Single crystals of TbMn$_2$O$_5$ were grown using
B$_2$O$_3$/PbO/PbF$_2$ flux in a Pt crucible. The flux was held at
1,280 $^{\circ}$C for 15 hours and slowly cooled down to 950
$^{\circ}$C at a rate of 1 $^{\circ}$C per hour. Crystals grew in
the form of black platelets as well as cubes with a typical size of
10 mm$^3$ with a working diameter of 5 mm. The crystals were
characterized and oriented using x-ray diffraction at room
temperature. Normal incidence reflection spectra were taken with a
Bomem Fourier Transform Spectrometer DA3.02. Light was polarized
along the \textit{a} and \textit{b} crystal axes in the frequency
range 8-800 cm$^{-1}$ ($\approx$ 1-100 meV) and in the temperature
range 7 to 300 K. Sample was kept in vacuum in a continuous He flow
cryostat with optical access windows. The factor group analysis
based on structural data by Alonso, et al \cite{Alonso-diffraction}
of the paraelectric phase gives the following IR active vibrational
modes at the $\Gamma$ point: $\Gamma_{IR} = 8B_{1u} (E||z) +
14B_{2u} (E||y) + 14B_{3u} (E||x)$ identical to a previous report
\cite{Mihailova-Raman}. We complemented this analysis with a shell
model calculation of the phonon dispersion.

\begingroup
\squeezetable
\begin{table}[h b]
\caption{Oscillator parameters at T$_1$ = 7 K and T$_2$ = 45 K in
TbMn$_2$O$_5$. \textit{a, b} are the crystal axes.
$\varepsilon_{\infty}^{a,b} = 5.31,6.82$. }
\begin{ruledtabular}
\begin{tabular}{cccc|cccc|cccc}
\multicolumn{4}{c|}{$\omega_o (cm^{-1})$} &
\multicolumn{4}{c|}{$\Delta \epsilon$} &
\multicolumn{4}{c}{$\gamma (cm^{-1})$}\\
\hline \multicolumn{2}{c}{\textit{a}} &
\multicolumn{2}{c|}{\textit{b}} & \multicolumn{2}{c}{\textit{a}} &
\multicolumn{2}{c|}{\textit{b}} &
\multicolumn{2}{c}{\textit{a}} & \multicolumn{2}{c}{\textit{b}} \\
T$_1$ & T$_2$  & T$_1$ & T$_2$ & T$_1$ & T$_2$  & T$_1$ & T$_2$ &
T$_1$ & T$_2$  & T$_1$ & T$_2$ \\
\hline
 111.9 & 111.7 & 97.2 & 96.4& 0.59 & 0.66 & 0.42 & 0.38 &
1.9 & 2 & 3.3 & 3.7\\
157.5 & 157.3 & 168.9 & 168.6& 0.81 & 1.05 & 0.46 & 0.43 &
0.9 & 0.3 & 1 & 1.1\\
164.2& 163.8 & 171.9 & 171.5& 1.68 & 2.02 & 0.30 & 0.35 &
3.3 & 2.6 & 1.4 & 1.5\\
218.5 & 217.4 & 222.2 & 221.9& 0.30 & 0.69 & 0.11 & 0.11 &
4.7 & 8.9 & 2.4 & 2.9\\
254.8 & 253.2 & 256.8 & 256.6& 1.88 & 2.56 & 0.17 & 0.18 &
3.1 & 1.5 & 2.1 & 2\\
333.1 & 332.4 & 333.4 & 332.7& 0.09 & 0.15 & 0.17 & 0.17 &
2.7 & 2.6 & 2.7 & 2.7\\
364.9 & 362.8 & 386 & 385.5 & 2.02 & 2.75 & 0.02& 0.01 &
3.9 & 1 &3.5& 4\\
397.6 & 396.5 & 422.3 & 422.3& 0.38 & 0.46 & 0.28 & 0.28 &
4.6 & 3.2 & 4 & 3.5 \\
494.8 & 493.9 &453.2 & 459.3 & 0.45 & 0.59 &3.43 & 3.56 &
5.3 & 3.7 &18.4 & 6.7 \\
613.5 & 611.3 & 481.8 & 483& 0.71 & 1.11 & 2.86 & 2.6 &
9 & 5.4 &4.4 & 3.3 \\
627.5 & 625.9 & 538.2 & 537.6& 0.23 & 0.14 &0.25 & 0.51  &
8.4 & 4.2 &7.3 & 7.1 \\
704.2 & 701.4 & 567.3 & 568.4& 0.05 & 0.04 &0.52 & 0.57  &
3.3 & 4.5 &5.1 & 7.9 \\
| & | &636.6 & 637.2 & | & | & 0.27 & 0.23 & | & | & 10.7 & 9.3\\
| & | &688.2 & 686.9 & | & | & 0.003 & 0.003 & | & | & 9.5 & 6\\
| & | &703\footnote{Previously IR inactive} & | & | & | & 0.001 & | & | & | & 7 & | \\
| & | &120.4\footnote{Crystal field excitation fitted as electric
dipole active} & 119.5 & | & | & 0.12 &
0.10 & | & | & 5.7 & 6.4 \\
\end{tabular}
\end{ruledtabular}
\label{Table-phonons}
\end{table}
\endgroup

\begin{figure}[h t]
\includegraphics[width=1\columnwidth]{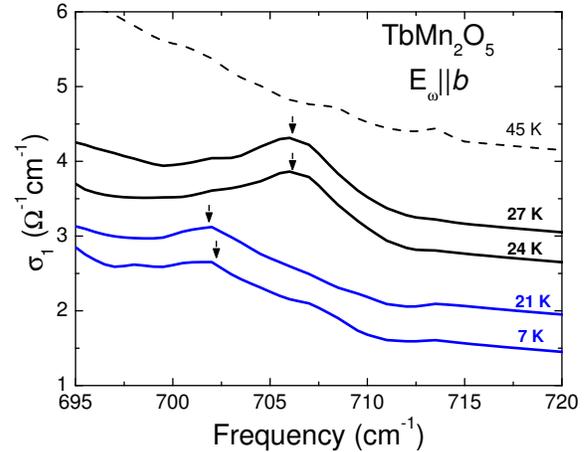}
\caption{(Color online). Optical conductivity of the newly
activated phonon (shifted plots). Arrows indicate the position of
the resonance frequency.} \label{commensuratephonon}
\end{figure}

Figure \ref{fullspectra} shows the spectra at T = 7 K. Twelve of
the 14 IR active phonons in the \textit{a} polarization were
reliably observed, whereas all 14 phonons polarized along
\textit{b} are present in the spectrum. The reflectivity spectra
was fitted in a least squares procedure using the sum of
Lorentzian form of the model dielectric function $\varepsilon$,
given by:
\begingroup
\begin{equation}
\varepsilon(\omega) = \varepsilon_{\infty} +
\sum_{i=1}^N{\frac{S_i}{\omega_{0
i}^{2}-\omega^{2}-\imath\gamma_{i}\omega}} \label{eps-lorentz}
\end{equation}
\endgroup
where $\varepsilon_{\infty}$ is the dielectric constant at high
frequency and the phonon parameters $\omega_{0}$, $S$ and $\gamma$
are the phonon frequency, the spectral weight and the linewidth
respectively, we also define $S_i = \Delta\epsilon_i \omega_{0
i}^{2}$ where $\Delta\epsilon$ is the contribution of the phonon to
the static dielectric function; these parameters are extracted as
functions of temperature and are displayed in table
\ref{Table-phonons}. The result of the fitting is also shown in
figure \ref{fullspectra} and it can be seen that it is almost
indistinguishable from the data indicating the weakness of higher
order phonon processes.

\begin{figure}
\includegraphics[width=1\columnwidth]{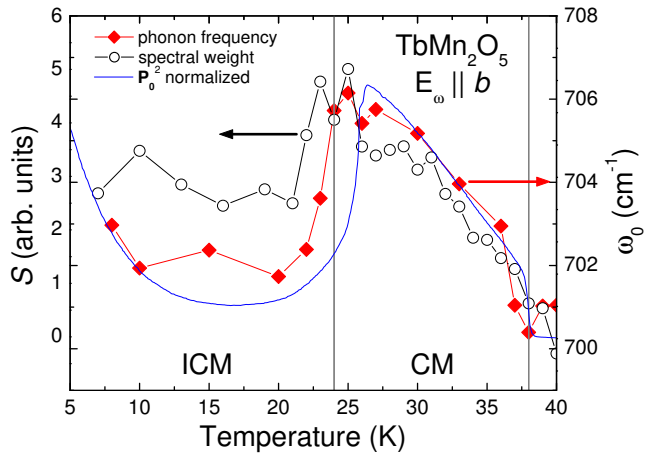}
\caption{(Color online). Comparison of the behavior of the spectral
weight and frequency of the new phonon to the static polarization
$\mathbf{P_0}$. Polarization and frequency show warming curves and
$S$ shows the cooling curve.} \label{Raman-phonon}
\end{figure}

We obtained the optical conductivity using the Kramers-Kronig
transform of the reflectivity spectrum. Figure
\ref{commensuratephonon} shows the optical conductivity around 700
cm$^{-1}$ for several temperatures with $E_\omega || \mathbf{P_0}$.
We observe that a feature not present at 45 K appears in the low T
phases. The temperature dependence of the spectral weight and
frequency of this feature are plotted in figure \ref{Raman-phonon}
where we see the spectral weight starting to appear at 38 K. $S$ for
this phonon was obtained by directly integrating the optical
conductivity between 695 and 710 cm$^{-1}$. The spectral weight of
this phonon increases and its frequency shifts, both continuously,
as we lower the temperature. Around 24 K both abruptly change and
show hysteresis around this point evident by the difference in the
cooling and warming curves. The behavior of this feature is
correlated with the second order FE transition at 38 K and the first
order CM $\rightarrow$ ICM transition at 24 K in this compound. The
static polarization $\mathbf{P_0}$ plotted in figure
\ref{Raman-phonon} was obtained by measuring the temperature
dependence of the pyroelectric current on a similar sample to the
one used in the optical measurements.

On very general grounds we can relate the appearance and behavior of
this phonon to these underlying phase transitions. Since the lattice
distortions $\delta u$ associated with these phase transitions are
very small we can expand the spectral weight and frequency shifts in
powers of $\delta u$. The quadratic term is the first non-zero term
in this expansion that can describe the spectral weight change or
frequency shift. Similarly, the order parameters associated with the
new phases are proportional to $\delta u$ so that $\mathbf{P_0}
\propto \delta u$. As a result we expect that the spectral weight
behavior and frequency shift should be $ S$, $\Delta\omega$
$\propto(\delta u)^2 \propto \mathbf{P_0}^2 $. This is the observed
behavior as can be seen in figure \ref{Raman-phonon} where we have
plotted $\mathbf{P_0}^2$ with the phonon data. At low temperatures
(T $<$ 10 K) where the Tb moment orders, the phonon data deviates
from $\mathbf{P_0}^2$ suggesting that it is the Mn and oxygen ion
displacements that dominate the dynamics of this high frequency
phonon. The possible scenarios for the appearance of a new phonon
are: (1) zone folding of the phonon dispersion (since the magnetic
order corresponds to a lock in ICM $\rightarrow$ CM transition with
$\mathbf{k} = (1/2,0,1/4)$), and (2) activation of IR-inactive
phonons at this transition due to the loss of inversion symmetry.
Our shell model calculation shows that the dispersion of the high
frequency phonon is negative so that no zone-folded mode can give
the high frequency of this phonon. We therefore conclude that this
phonon is a previously IR inactive phonon that acquires electric
dipole moment at the FE transition.

\begingroup
\begin{table}
\caption{Irreducible representation splitting at the FE phase
transition. S. A. = Spectral Activity (R = Raman active, IR =
Infrared Active).}
\begin{ruledtabular}
\begin{tabular}{c|c|c|c}
\# 55 S. A. & \# 55 irreps & \# 26 irreps & \#26 S. A.\footnote{Note
that in this column x, y, z correspond to the high temperature
system of coordinates \textit{a, b, c} and
differ from what is found in the character table.}\\
\hline
R &$A_g$ & $A_1$ & IR (y) \& R\\
R &$B_{1g}$ & $B_2$ & IR (x) \& R\\
R & $B_{2g}$ & $A_2$ & R\\
R &$B_{3g}$ & $B_1$ & IR (z) \& R\\
Silent &$A_u$ & $A_2$ & R\\
IR(z) &$B_{1u}$ & $B_1$ & IR (z) \& R\\
IR(y) &$B_{2u}$ & $A_1$ & IR (y) \& R\\
IR(x) &$B_{3u}$ & $B_2$ & IR (x) \& R\\
\end{tabular}
\end{ruledtabular}
\label{Table-irrep}
\end{table}
\endgroup
In a ferroelectric phase transition, where inversion symmetry is
lost, symmetry considerations dictate that phonons that were not IR
active in the paraelectric phase can become IR active in the FE
phase. This is the case in TbMn$_2$O$_5$, where the low T phase has
mixed IR and Raman phonons. In this low T phase the phonons of the
high T symmetry group split as shown in table \ref{Table-irrep}
\cite{Symmetry-note}. This splitting was obtained by considering
what symmetry operation is maintained in both phases and then assign
spectral activity according to the experimental observations. The
assumption was made as well that the low T space group is as
proposed by Polyakov, et al \cite{Polyakov-neutron} and Kagomiya, et
al \cite{Kagomiya-xray}. Consistency with the laboratory frame (x
$\rightarrow a$, y $\rightarrow b$, z $\rightarrow c$) is applied as
well. From the reports by Mihailova, et al. \cite{Mihailova-Raman}
and Garc\'{i}a-Flores, et al. \cite{Garcia-Flores-Raman} we learn
that a $A_{g}$ mode at frequency $\approx$ 700 cm$^{-1}$ exists in
all the \emph{R}Mn$_2$O$_5$ materials whose Raman phonons have been
reported. We note as well that these reports do not resolve any IR
phonons becoming Raman active at the FE phase transition. We
conclude that this high frequency $A_{g}$ Raman phonon is the mode
we observe that acquires IR activity in the FE phase. Further
experimental confirmation in the Raman spectrum is needed to make a
definitive identification.

\begin{figure}
\includegraphics[width=1\columnwidth]{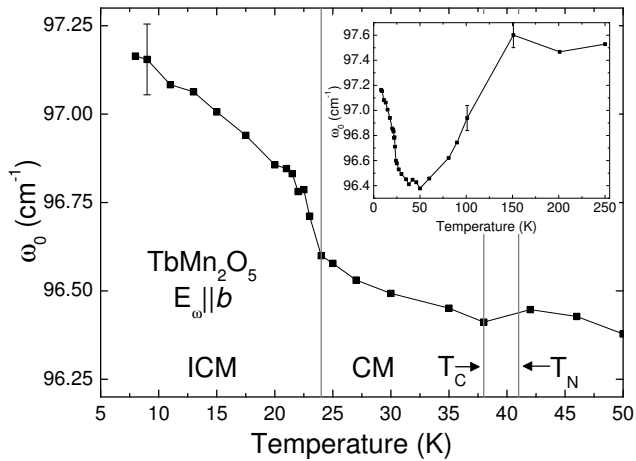}
\caption{Temperature dependence of the \emph{b} axis Tb phonon
frequency. Inset shows the full temperature dependence. Typical
dispersion of several measurements is indicated as error
bars.}\label{tbphonon}
\end{figure}

Only a few other phonons show correlations with the low temperature
phase transitions.  The phonons polarized along the \textit{a} axis
do not show any significant anomalies in this temperature range.
This is consistent with the featureless behavior of the dielectric
function along this axis. On the other hand, the behavior of some of
the phonons with dipole moment along \textit{b} is non-trivial. The
low frequency phonon with frequency $\approx$ 96 cm$^{-1}$,
identified primarily with movement of the Tb ions, has a temperature
dependence that correlates with the low temperature CM $\rightarrow$
ICM magnetic transition. In figure \ref{tbphonon} the frequency of
this phonon is plotted versus temperature and we observe an increase
in the frequency around 24 K. This effect is thought to be a
manifestation of the coupling of this phonon to a magnon as is
discussed by Katsura, et al \cite{Katsura-electromagnon}. Further
experimental results on this effect will be presented elsewhere
\cite {Andrei-electromagnon}.

\begin{figure}
\includegraphics[width=1\columnwidth]{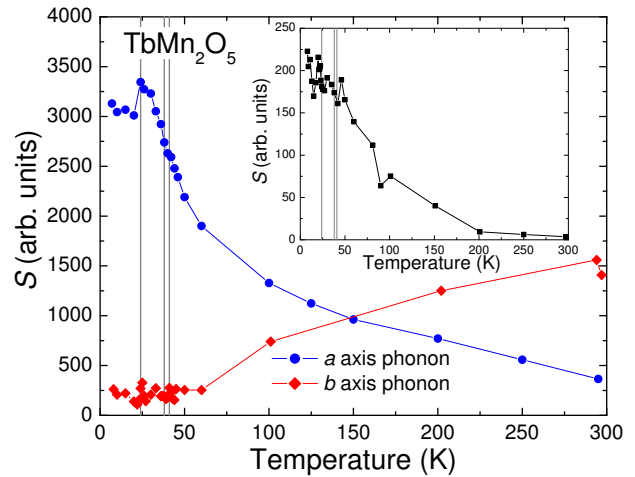}
\caption{(Color online). Temperature dependence of the spectral
weight of 2 oxygen dominated phonons. Inset shows the temperature
dependence of the spectral weight of one of the Tb$^{3+}$ crystal
field transitions.} \label{corrphonons}
\end{figure}

Surprisingly several phonons show interesting temperature
dependence for T above T$_N$. The inset in figure \ref{tbphonon}
shows the full temperature dependence of the frequency of the
\textit{b} axis phonon. The anomalous softening in the temperature
range of 150 K to 50 K demonstrates additional effects in the
dynamics of the lattice. In figure \ref{corrphonons} the behavior
of the spectral weight of two oxygen phonons polarized in the
\emph{a} and \emph{b} axes, with frequencies of 704 and 689
cm$^{-1}$ respectively, seems complementary: the \emph{a} phonon
gains spectral weight while the \emph{b} phonon looses it as the
temperature is lowered. This effect is present in the full
temperature range from 300 K to 7 K, while for the rest of the
phonons the spectral weight is only changed significantly around
the various transition temperatures. The fact that these modes
gain or lose so much spectral weight (10 and 6-fold respectively)
in a large temperature interval also demonstrates some higher
energy scale in this system. These effects (the anomalous
softening and the dramatic changes in spectral weight) are not
understood at present. However, one interesting possibility is
that they are of magnetic character. Recent high temperature
susceptibility measurements \cite{Garcia-Flores-Raman} in
BiMn$_2$O$_5$ has shown evidence for spin frustrated behavior from
deviations from Curie law with a Weiss temperature of $\approx$
250 K. Dielectric anomalies in BiMn$_2$O$_5$ around this
temperature have been reported \cite{Golovenchits-high-T} as well.

Finally the inset of figure \ref{corrphonons} shows the temperature
dependence of the intensity of a feature observed at 120 cm$^{-1}$.
This is identified as a crystal field level of the Tb $^{3+}$ ion
\cite{Gingras-Tb-CFT}. This transition has electric dipole character
as is seen from the form of the reflectivity curve (see fig.
\ref{fullspectra}) as well as the fact that the spectral weight (see
table \ref{Table-phonons}) is comparable to the IR active phonons
(magnetic dipole transitions are usually much weaker than electric
dipole transitions). This conclusion is supported as well by the
shell model calculation that shows the 3 lowest phonon excitations
being the Tb-dominated phonon (at $\approx$ 100 cm$^{-1}$) and then
a doublet (at $\approx$ 170 cm$^{-1}$). Furthermore, the observed
temperature dependence of the intensity is common for the f-level
crystal field transitions in the rare earth ions
\cite{Jandl-crystal}. Several other absorption features are observed
in transmission at lower frequencies and will be reported elsewhere
\cite{Andrei-electromagnon}.

\section{Conclusions}
We have measured the IR phonon spectra in TbMn$_2$O$_5$ along the
\textit{a} and \textit{b} axes and have observed most of the
symmetry allowed modes. The majority of the phonons do not show
significant correlations to the FE and AFM phase transitions of
the system. However several phonons exhibit interesting
correlations to the ferroelectricity of this material. We have
found a signature of the loss of inversion symmetry in the FE
phase by the appearance of  a IR phonon below $T_c$ that was only
Raman active in the paraelectric phase. The strength of this mode
is proportional to the square of the FE order parameter and gives
a sensitive measure of the symmetry lowering lattice distortions
in the ferroelectric phase. In addition, the lowest frequency
phonon (along \textit{b}) displays hardening in the CM
$\rightarrow$ ICM transition possibly due to the coupling with the
spin system \cite{Katsura-electromagnon}; 2 modes (along
\textit{a} and \textit{b}) have dramatic changes in their spectral
weight over a wide temperature range possibly because of
frustration effects in the spin system. We have also identified an
electric dipole active crystal field transition of the Tb$^{3+}$
ion in the phonon frequency range.
\section{Acknowledgements}
We thank G-W. Chern, D.I. Khomskii, S. Jandl, J. Simpson and O.
Tchernyshyov for useful discussions. This work was supported by
the National Science Foundation MRSEC under Grant No. DMR-0520471.

\bibliography{phonons-TbMn2O5}
\end{document}